\documentclass[12pt]{article}
\usepackage{amsmath,amssymb,amsfonts,amsthm}

\title{Entanglement dynamics and position-momentum entropic uncertainty relation of a $\Lambda$-type three-level atom interacting with a two-mode cavity field in the presence of nonlinearities}

\author{M J Faghihi$^{1,4,5}$, M K Tavassoly$^{1,2,4,*}$ and M R Hooshmandasl$^{3,4}$ \\
 \footnotesize{$^1$ Atomic and Molecular Group, Faculty of Physics, Yazd University, Yazd, Iran} \\
 \footnotesize{$^2$ Photonics Research Group, Engineering Research Center, Yazd University, Yazd, Iran} \\
 \footnotesize{$^3$ Department of Computer Science, Yazd University, Yazd, Iran} \\
 \footnotesize{$^4$ The Laboratory of Quantum Information Processing, Yazd University, Yazd, Iran} \\
 \footnotesize{$^5$ Physics and Photonics Department, Graduate University of Advanced Technology, Mahan, Kerman, Iran} \\
 \footnotesize{$^*$ E-mail: mktavassoly@yazd.ac.ir}}

\begin{document}
\maketitle
 \newcommand{\norm}[1]{\left\Vert#1\right\Vert}
 \newcommand{\abs}[1]{\left\vert#1\right\vert}
 \newcommand{\set}[1]{\left\{#1\right\}}
 \newcommand{\R}{\mathbb R}
 \newcommand{\I}{\mathbb{I}}
 \newcommand{\C}{\mathbb C}
 \newcommand{\eps}{\varepsilon}
 \newcommand{\To}{\longrightarrow}
 \newcommand{\BX}{\mathbf{B}(X)}
 \newcommand{\HH}{\mathfrak{H}}
 \newcommand{\A}{\mathcal{A}}
 \newcommand{\D}{\mathcal{D}}
 \newcommand{\N}{\mathcal{N}}
 \newcommand{\x}{\mathcal{x}}
 \newcommand{\p}{\mathcal{p}}
 \newcommand{\la}{\lambda}
 \newcommand{\af}{a^{ }_F}
 \newcommand{\afd}{a^\dag_F}
 \newcommand{\afy}{a^{ }_{F^{-1}}}
 \newcommand{\afdy}{a^\dag_{F^{-1}}}
 \newcommand{\fn}{\phi^{ }_n}
 \newcommand{\HD}{\hat{\mathcal{H}}}
 \newcommand{\HDD}{\mathcal{H}}

 \begin{abstract}
 In this paper, the interaction between a $\Lambda$-type three-level atom and two-mode cavity field is discussed. The detuning parameters and cross-Kerr nonlinearity are taken into account and it is assumed that atom-field coupling and Kerr medium to be $f$-deformed. Even though the system seems to be complicated, the analytical form of the state vector of the entire system for considered model is exactly obtained. The time evolution of nonclassical properties such as quantum entanglement and position-momentum entropic uncertainty relation (entropy squeezing) of the field are investigated. In each case, the influences of the detuning parameters, generalized Kerr medium and intensity-dependent coupling on the latter nonclassicality signs are analyzed, in detail.
 \end{abstract}

 \section{Introduction}\label{sec-intro}
 Quantum mechanical point of view of atom-field interaction, most of the time, shows new features of quantum nature of the field as well as the atom. The quantum description of atom-field interaction is elegantly described by the well-known Jaynes-Cummings model (JCM) \cite{JCM}. This model is a simplified version of the interaction of the quantized (single-mode) electromagnetic field and the two-level atom in the rotating wave approximation (RWA). Many various generalizations have been proposed to modify the JCM in the literatures. Intensity-dependent (nonlinear) JCM was suggested by Buck and Sukumar \cite{suk} and then used by others \cite{buzek,selective}. Fang {\it et al} examined the entropy and phase properties of the field in two-photon JCM at the influence of the nonlinear interaction of a Kerr medium with the field mode \cite{mao}. Kazakov discussed the interaction of two-level atom with two-mode field \cite{kazakov}. Crnugelj {\it et al} studied the time evolution and squeezing properties of the deformed JCM which corresponds to the usual model with intensity-dependent coupling controlled by two additional parameters may be determined by experiment \cite{crn}. Abdel-Wahab {\it et al} discussed about the interaction between the two two-level atoms and two-mode field and investigated the effect of the detuning parameters and cross-Kerr nonlinearity on some of the nonclassical properties of the considered system \cite{wahab}. Quantum properties of a $V$-type and a $\Lambda$-type three-level atom interacting with a single-mode field in a Kerr medium with intensity-dependent coupling and in the presence of the detuning parameters have been studied in \cite{zait} and \cite{us} by Zait and us, respectively. Guo {\it et al} \cite{guo} examined individually the influences of the Kerr-like medium and intensity-dependent coupling on entropy exchange and entanglement in the JCM and showed that these effects are helpful for improving the quality of entropy exchange. A generalization of the time-dependent JCM including nonlinear terms in the field and/or an intensity-dependent coupling term has been considered by Cordero {\it et al} \cite{rec1}, in which they studied the effect of atomic motion in the cavity on its quantum state. Using the framework of $f$-deformed algebra, nonlinear JCM in the presence of a Kerr-like medium has been introduced by S$\acute{\mathrm{a}}$nchez {\it et al} \cite{rec2} where they studied some quantum properties of the nonlinear field.  A general formalism for a $\Lambda$-type three-level atom interacting with a correlated two-mode field is presented by Abdel-Aty {\it et al} \cite{aty}.  The authors found the degree of entanglement (DEM) for their system by obtaining the density matrix operator.\\
 In this paper, we intend to suggest a model that describes the interaction between a three-level atom (which is considered in a $\Lambda$-configuration) and a two-mode cavity field in the presence of various nonlinearities consist of a Kerr and deformed Kerr medium (which is distinguished from a Kerr medium by a nonlinearity function that depends on the intensity of light), intensity-dependent atom-field coupling and the detuning parameters.
 Apart from other new features of our work, in particular, we investigate the effect of deformed Kerr medium on the atom-field interaction, in the presence of
 intensity-dependent atom-field coupling and the detuning parameters.
 The main goal of the present paper is to discuss on the effects of these parameters on the physical properties of the state vector of the whole system. To achieve this purpose, we firstly evaluate the time evolution of the field entropy by which  the amount of entanglement between subsystems is determined. In addition, the position-momentum entropic uncertainty relation is considered, from which the entropy squeezing can be investigated.\\
 The remainder of paper is organized as follows: In the next section, we obtain the state vector of the whole system using the generalized JCM. Then, in section 3, the DEM between the atom and the field is evaluated and then, entropy squeezing is studied in section 4. Finally, section 5 contains a summary and concluding remarks.

 \section{The model}

 This section is allocated to the description of the whole system under our considerations.
 We concentrate on the interaction between a three-level atom and a two-mode cavity field involving deformed Kerr medium with intensity-dependent atom-field coupling and the detuning parameters. Let us consider a model in which the two-mode quantized electromagnetic field oscillates with frequencies $\Omega_{1}$ and $\Omega_{2}$ in an optical cavity and the three-level atom be in the $\Lambda$-type atomic configuration. In this atomic configuration with atomic levels indicated by $|j\rangle$ and energies $\omega_{j}$, where j=1,2,3 (see figure 1), the transitions $|1\rangle\rightarrow|2\rangle$ and $|1\rangle\rightarrow|3\rangle$ are allowed and the transition $|2\rangle\rightarrow|3\rangle$ is forbidden in the electric-dipole approximation \cite{zubairy}. According to the generalized JCM, the Hamiltonian for such system in the RWA can be written as ($\hbar=c=1$):
 \begin{eqnarray}
 \hat{H} =\hat{H}_{0}+\hat{H}_{1}, \nonumber
 \end{eqnarray}
 where
 \begin{eqnarray}\label{hamiltonih0}
 \hat{H}_{0}&=& \sum_{i=1}^{3} \omega_{i}\hat{\sigma}_{ii}+\sum_{j=1}^{2} \Omega_{j} \hat{a}^{\dag}_{j} \hat{a}_{j},
 \end{eqnarray}
 and
 \begin{eqnarray}\label{hamiltonih1}
 \hat{H}_{1}= \chi \hat{R}_{1}^{\dag} \hat{R}_{1} \hat{R}_{2}^{\dag} \hat{R}_{2}+\lambda_{1}(\hat{\A}_{1} \hat{\sigma}_{12}+\hat{\sigma}_{21}\hat{\A}_{1}^{\dag})
 + \lambda_{2}(\hat{\A}_{2}\hat{\sigma}_{13}+\hat{\sigma}_{31}\hat{\A}_{2}^{\dag})
 \end{eqnarray}
 where $\hat{\sigma}_{ij}$ is the atomic lowering and raising operator between $|i\rangle$ and $|j\rangle$ defined by $\hat{\sigma}_{ij}=|i\rangle \langle j|,(i,j=1,2,3),\hat{a}_{j}$ ($\hat{a}_{j}^{\dag}$) is the bosonic annihilation (creation) operator of the field mode $j$, $\chi $ denotes the nonlinearity of the deformed cross-Kerr medium and the constants $\lambda_{1},\lambda_{2}$ determine the strength of the atom-field couplings. In above relation, $\hat{R}_{j}=\hat{a}_{j}\hat{g}_{j}(\hat{n}_{j})$ and $\hat{\A}_{j}=\hat{a}_{j}\hat{f}_{j}(\hat{n}_{j})$, and $\hat{R}_{j}^{\dag}$ and $\hat{\A}_{j}^{\dag}$ are their respective conjugate hermitians, where $\hat{n}_{j}=\hat{a}_{j}^{\dag}\hat{a}_{j}$. Also, $g_{j}(\hat{n}_{j})$ and $f_{j}(\hat{n}_{j})$ are two generally different operator-valued functions which determine the deformation of the Kerr medium and intensity-dependent atom-field coupling, respectively. Therefore, we deal with two sets of operators satisfying the following commutation relations:
 \begin{eqnarray}\label{vrrd2}
 [\hat{R}_{j},\hat{n}_{j}]=\hat{R}_{j}, \;\;\;\;\;\;\;\;\; [\hat{R}_{j}^{\dag},\hat{n}_{j}]=-\hat{R}_{j}^{\dag},\;\;\;\;\;\;\;\;\;[\hat{R}_{i},\hat{R}_{j}]=0,
  \end{eqnarray}
 and
 \begin{eqnarray}\label{vrrd21}
 [\hat{\A}_{j} ,\hat{n}_{j}]=\hat{\A}_{j}, \;\;\;\;\;\;\;\;\; [\hat{\A}_{j}^{\dag},\hat{n}_{j}]=-\hat{\A}_{j}^{\dag}, \;\;\;\;\;\;\;\;\; [\hat{\A}_{i},\hat{\A}_{j}]=0,
  \end{eqnarray}
 where we assumed $g(\hat{n}_{j})$ and $f(\hat{n}_{j})$ to be real functions of number operators.
 A deep insight in the form of the introduced Hamiltonian $\hat{H}_{1}$ in (\ref{hamiltonih1}) and comparing with full linear regime, shows that
  the constant susceptibility $\chi$ and coupling constants $\lambda_{i}$ ($i=1,2$) are changed to deformed $\chi$, $\chi g^{2}_{1}(n_{1}) g^{2}_{2}(n_{2})$, (the physical motivation of this change has been established in \cite{newhonarasa}) and intensity-dependent coupling, $\lambda_{i} f_{i}(n_{i})$, respectively.
 It is noteworthy to mention that in the continuation of this paper, we will select the nonlinearity functions as $f_{i}(n_{i})=\sqrt{n_{i}}$ and $g_{i}(n_{i})=1/\sqrt{n_{i}}$. By these considerations, one can rewrite the explicit form of the interaction Hamiltonian in the final form
  \begin{eqnarray}\label{hamiltonih1final}
 \hat{H}_{1}^{\mathrm{int}} = \chi +\lambda_{1} (\hat{a}_{1} \sqrt{\hat{a}_{1}^{\dag}\hat{a}_{1}} \hat{\sigma}_{12} +
 \hat{\sigma}_{21} \sqrt{\hat{a}_{1}^{\dag}\hat{a}_{1}}\hat{a}_{1}^{\dag})
 + \lambda_{2}(\hat{a}_{2} \sqrt{\hat{a}_{2}^{\dag}\hat{a}_{2}} \hat{\sigma}_{13}+\hat{\sigma}_{31}\sqrt{\hat{a}_{2}^{\dag}\hat{a}_{2}}\hat{a}_{2}^{\dag}).
 \end{eqnarray}
 Nevertheless, in order to obtain a general formalism for our considered system, we proceed our calculations with general form of  $f_{i}(n_{i})$ and $g_{i}(n_{i})$.\\
 Anyway, let us now consider the wave function $|\psi(t)\rangle$ corresponding to the whole system at any time $t$ to be in the form
 \begin{eqnarray}\label{say}
 |\psi(t)\rangle&=&\sum_{n_{1}=0}^{+\infty}\sum_{n_{2}=0}^{+\infty}q_{n_{1}}q_{n_{2}}\Big[ A(n_{1},n_{2},t) e^{-i\gamma_{1} t}|1,n_{1},n_{2}\rangle \nonumber \\
 &+&B(n_{1}+1,n_{2},t)e^{-i\gamma_{2} t}|2,n_{1}+1,n_{2}\rangle
 + C(n_{1},n_{2}+1,t)e^{-i\gamma_{3} t}|3,n_{1},n_{2}+1\rangle  \Big],
 \end{eqnarray}
 where $q_{n_{1}}$ and $q_{n_{2}}$ describe the amplitudes of the initial field states, $A,B$ and $C$ are the time-dependent atomic probability amplitudes which have to be evaluated and
 \begin{eqnarray}
 \gamma_{1}&=&\omega_{1}+n_{1}\Omega_{1}+n_{2}\Omega_{2}, \nonumber \\
 \gamma_{2}&=&\omega_{2}+(n_{1}+1)\Omega_{1}+n_{2}\Omega_{2}, \nonumber \\
 \gamma_{3}&=&\omega_{3}+n_{1}\Omega_{1}+(n_{2}+1)\Omega_{2}.
 \end{eqnarray}
 At this stage, we should find the atomic probability amplitudes which determine the explicit form of the wave function of whole system. Putting the wave function introduced in (\ref{say}) into the time-dependent Schr\"{o}dinger equation
 \begin{eqnarray}\label{sh-eq}
  i \frac{\partial}{\partial t} |\psi(t)\rangle =\hat{H}| \psi(t)\rangle
 \end{eqnarray}
 arrives one at the following coupled differential equations for the atomic probability amplitudes
 \begin{eqnarray}\label{coupling}
 i\;\dot{A}&=& V_{A} A + \kappa_{1}B e^{-i\Delta_{2}t}+ \kappa_{2} Ce^{-i\Delta_{3}t}, \nonumber \\
 i\;\dot{B}&=&V_{B} B+\kappa_{1} A e^{i\Delta_{2}t}, \nonumber \\
 i\;\dot{C}&=&V_{C} C+ \kappa_{2} A e^{i\Delta_{3}t},
 \end{eqnarray}
 where the detuning parameters $\Delta_{2}$ and $\Delta_{3}$ are now given by
 \begin{eqnarray}\label{cavp}
 \Delta_{2}=\omega_{2}-\omega_{1}+\Omega_{1}, \;\;\;\;\; \Delta_{3}=\omega_{3}-\omega_{1}+\Omega_{2},
 \end{eqnarray}
 and the dot signs refer to the time differentiation. In (\ref{coupling}) we have set
 \begin{eqnarray}\label{vdefinition1}
 V_{A}=V(n_{1},n_{2}), \;\;\;\;\; V_{B}=V(n_{1}+1,n_{2}),\;\;\;\;\; V_{C}=V(n_{1},n_{2}+1), \nonumber \\
 \kappa_{1}=\lambda_{1}\;\sqrt{n_{1}+1}\;f_{1}(n_{1}+1), \;\;\;\;\; \kappa_{2}=\lambda_{2}\;\sqrt{n_{2}+1}\;f_{2}(n_{2}+1),
 \end{eqnarray}
 with
 \begin{eqnarray}\label{vdefinition2}
 V(n_{1},n_{2})=\chi\; n_{1} n_{2} g_{1}^{2}(n_{1})g_{2}^{2}(n_{2}).
 \end{eqnarray}
 By inserting $B=e^{i\mu t}$ in equation ({\ref{coupling}}) we obtain the following relation
 \begin{eqnarray}\label{mu}
 \mu ^{3}+x_{1} \mu ^{2}+x_{2} \mu + x_{3}=0,
 \end{eqnarray}
 where
 \begin{eqnarray}\label{x123}
 x_{1}&=&V_{A}+V_{B}+V_{C}+\Delta_{3}-2\Delta_{2}, \nonumber \\
 x_{2}&=& (V_{A}+V_{B}-\Delta_{2})(V_{C}+\Delta_{3}-\Delta_{2})+V_{B}(V_{A}-\Delta_{2})-\kappa_{1}^{2}-\kappa_{2}^{2}, \nonumber \\
 x_{3}&=&V_{B}\left[(V_{A}-\Delta_{2})(V_{C}+\Delta_{3}-\Delta_{2})-\kappa_{2}^{2} \right]-\kappa_{1}^{2}(V_{C}+\Delta_{3}-\Delta_{2}).
 \end{eqnarray}
 It is clear that equation (\ref{mu}) has generally three different roots. So, $B$ will be a linear combination of $e^{i\mu_{j}t}$ as follows
 \begin{eqnarray}\label{b}
 B(t)=\sum_{j=1}^{3} \tilde{b_{j}}e^{i\mu_{j} t},\;\;\;\;\tilde{b_{j}}=\kappa_{1}b_{j}.
 \end{eqnarray}
 The general solution for equation (\ref{mu}) is given by
 \begin{eqnarray}\label{vkardan}
 \mu_{j}&=&-\frac{1}{3}x_{1}+\frac{2}{3}\sqrt{x_{1}^{2}-3x_{2}}\cos\left[ \theta+\frac{2}{3}(j-1)\pi \right],\;\;\;\;\;j=1,2,3, \nonumber \\
 \theta &=& \frac{1}{3}\cos^{-1}\left[ \frac{9x_{1}x_{2}-2x_{1}^{3}-27x_{3}}{2(x_{1}^{2}-3x_{2})^{3/2}}\right],
 \end{eqnarray}
 where we have used Kardan's formula \cite{kardan}. Now, by replacing equation (\ref{b}) into the coupled differential equations in (\ref{coupling})
 and after some lengthy but straightforward manipulations we obtain the probability amplitudes in the following form:
 \begin{eqnarray}\label{abc}
 A(n_{1},n_{2},t)&=&-e^{-i\Delta_{2}t}\sum_{j=1}^{3}(\mu_{j}+V_{B})b_{j}e^{i\mu_{j}t}, \nonumber \\
 B(n_{1}+1,n_{2},t)&=&\sum_{j=1}^{3}\kappa_{1}\;b_{j} e^{i\mu_{j} t}, \nonumber \\
 C(n_{1},n_{2}+1,t)&=&\frac{1}{\kappa_{2}}e^{i(\Delta_{3}-\Delta_{2})t}\sum_{j=1}^{3}\Big[(\mu_{j}+V_{B})
 (\mu_{j}+V_{A}-\Delta_{2})-\kappa_{1}^{2}\Big]b_{j}e^{i\mu_{j}t},
 \end{eqnarray}
 where the coefficients $b_{j}$ are still unknown parameters but they can be determined by specifying the initial condition of atom. Hence, by considering the atom to be initially in the excited state, i.e. $A(0)=1$, $B(0)=C(0)=0$ and using the equations in (\ref{abc}), the following relations for $b_{j}$  may be found
 \begin{eqnarray}\label{b123}
 b_{j}=\frac{\mu_{k}+\mu_{l}+V_{A}+V_{B}-\Delta_{2}}{\mu _{jk} \mu _{jl}},\;\;\;\;\;\;j\neq k\neq l=1,2,3,
 \end{eqnarray}
 where $ \mu _{jk}=\mu_{j}-\mu_{k} $. Summing up, the wave function $|\psi(t)\rangle$ as given in (\ref{say}) is exactly obtained.
 Altogether, it should be emphasized that the time-dependent atomic probability amplitudes $A$, $B$ and $C$ in the state vector of the system obtained in closed form in (\ref{abc}) are very complicated. The dependence of the coefficients on the nonlinearity functions $g_{j}(n_{j})$ and $f_{j}(n_{j})$ may be observed by an inverse overview on the relations (\ref{b123}) to (\ref{vdefinition1}). Anyway, now we are able to study the nonclassical properties of the state of the atom-field system.

 \section{Quantum mutual information and the DEM}
  Quantum entanglement is one of the most essential characteristics of the quantum mechanical systems which plays a key role within new information technologies and in many of the interesting applications of quantum computation and quantum information \cite{benenti}. Since quantum entanglement is considered as a basic ingredient in the vivid description of the structural properties of composite quantum systems, the analysis of the entanglement properties of subsystems is of more interest. In order to study the entanglement dynamics quantitatively and to evaluate the DEM, the quantum entropy (quantum mutual information) is a useful criterion that indicates the amount of entanglement \cite{chuang}. In other words, the time evolution of the entropy of the field or the atom reflects the time evolution of the DEM between subsystems. In the present bipartite quantum system under our consideration (atom and field), we use the von Neumann entropy \cite{pk1}. Before obtaining the reduced entropy of the field and the atom, it is valuable to pay attention to the important theorem of Araki and Leib \cite{araki}. According to this theorem, in a bipartite quantum system, the system and subsystem entropies, at any time $t$, are limited by the following triangle inequality
 \begin{eqnarray}\label{valen}
 |S_{A}(t)-S_{F}(t)|\leq S_{AF}(t) \leq S_{A}(t)+S_{F}(t),
 \end{eqnarray}
 where here the subscripts $` `$A$"$ and $` `$F$"$ refer to the atom and the field, respectively and the total entropy of the atom-field
 system is denoted by $S_{AF}$. As a result, if at the initial time the field and the atom are in pure states, the total entropy of the system is zero and remains constant. This means that, if the system is initially prepared in a pure state (as we have considered), at any time $t>0$, the reduced entropies of the two subsystems (atom and field) are identical, that is, $S_{A}(t)=S_{F}(t)$ \cite{phoenix}. So, we need only to calculate the reduced entropy of the atom to arrive at the DEM.
 According to the von Neumann entropy, as a measure of entanglement and at the same time, the degree of purity loss of the subsystem, the entropy of the atom and the field are defined through the corresponding reduced density operator by
 \begin{eqnarray}\label{ventd}
 S_{A(F)}(t)=-\mathrm{Tr}_{A(F)} \left(\hat{\rho}_{A(F)}(t) \ln \hat{\rho}_{A(F)}(t) \right).
 \end{eqnarray}
 The reduced density operator of the atom is given by
 \begin{eqnarray}\label{vrdma}
 \hat{\rho}_{A}(t)&=&\mathrm{Tr}_{F}\left(  |\psi(t) \rangle \langle \psi(t) |   \right)  \nonumber \\
 &=&\left(
  \begin{array}{ccc}
    \rho_{11} & \rho_{12} & \rho_{13} \\
    \rho_{21} & \rho_{22} & \rho_{23} \\
    \rho_{31} & \rho_{32} & \rho_{33} \\
  \end{array}
 \right),
 \end{eqnarray}
 where the elements of symmetric matrix in (\ref{vrdma}) are given below:
 \begin{eqnarray}\label{rho}
 \rho_{11}&=&\sum_{n_{1}=0}^{+\infty} \sum_{n_{2}=0}^{+\infty} q_{n_{1}} q^{*}_{n_{1}} q_{n_{2}} q^{*}_{n_{2}} A(n_{1},n_{2},t) A^{*}(n_{1},n_{2},t), \nonumber \\
 \rho_{12}&=&\sum_{n_{1}=0}^{+\infty} \sum_{n_{2}=0}^{+\infty} q_{n_{1}+1} q^{*}_{n_{1}} q_{n_{2}} q^{*}_{n_{2}} A(n_{1}+1,n_{2},t) B^{*}(n_{1}+1,n_{2},t) \exp(i\Delta_{2} t) = \rho^{*}_{21}, \nonumber \\
 \rho_{13}&=&\sum_{n_{1}=0}^{+\infty} \sum_{n_{2}=0}^{+\infty} q_{n_{1}} q^{*}_{n_{1}} q_{n_{2}+1} q^{*}_{n_{2}} A(n_{1},n_{2}+1,t) C^{*}(n_{1},n_{2}+1,t) \exp(i\Delta_{3} t) =\rho^{*}_{31}, \nonumber \\
 \rho_{22}&=&\sum_{n_{1}=0}^{+\infty} \sum_{n_{2}=0}^{+\infty} q_{n_{1}} q^{*}_{n_{1}} q_{n_{2}} q^{*}_{n_{2}} B(n_{1}+1,n_{2},t) B^{*}(n_{1}+1,n_{2},t), \nonumber \\
 \rho_{23}&=&\sum_{n_{1}=0}^{+\infty} \sum_{n_{2}=0}^{+\infty} q_{n_{1}} q^{*}_{n_{1}+1} q_{n_{2}+1} q^{*}_{n_{2}} B(n_{1}+1,n_{2}+1,t) C^{*}(n_{1}+1,n_{2}+1,t) \exp(i(\Delta_{3}-\Delta_{2}) t) =\rho^{*}_{32}, \nonumber \\
 \rho_{33}&=&\sum_{n_{1}=0}^{+\infty} \sum_{n_{2}=0}^{+\infty} q_{n_{1}} q^{*}_{n_{1}} q_{n_{2}} q^{*}_{n_{2}} C(n_{1},n_{2}+1,t) C^{*}(n_{1},n_{2}+1,t),
 \end{eqnarray}
 in which, in all of the above relations, $q_{n_{1}}$ and $q_{n_{2}}$ are the probability amplitudes of the initial radiation field modes, and $A$, $B$, $C$ are the
 atomic probability amplitudes have been introduced in (\ref{abc}). Consequently, the entropy of the field or the atom can be obtained by the following
 relation \cite{us,chuang,pk3}
 \begin{eqnarray}\label{sff}
 \mathrm{DEM}(t)=S_{F}(t)=S_{A}(t)=-\sum_{j=1}^{3}\xi_{j} \ln \xi_{j}
 \end{eqnarray}
 where $\xi_{j}$ are the eigenvalues of the reduced atomic density operator in (\ref{vrdma}) which have been given by Kardan's instruction as \cite{kardan}
 \begin{eqnarray}\label{ventkardan}
 \xi_{j}&=&-\frac{1}{3}\alpha_{1}+\frac{2}{3}\sqrt{\alpha_{1}^{2}-3\alpha_{2}}\cos\left[\beta+\frac{2}{3}(j-1)\pi \right], \nonumber \\
 \beta &=& \frac{1}{3}\cos^{-1}\left[ \frac{9\alpha_{1}\alpha_{2}-2\alpha_{1}^{3}-27\alpha_{3}}{2(\alpha_{1}^{2}-3\alpha_{2})^{3/2}}\right],
 \end{eqnarray}
 with
 \begin{eqnarray}\label{vzal}
 \alpha_{1}&=&-\rho_{11}-\rho_{22}-\rho_{33}, \nonumber \\
 \alpha_{2}&=&\rho_{11}\rho_{22}+\rho_{22}\rho_{33}+\rho_{33}\rho_{11} -\rho_{12}\rho_{21}-\rho_{23}\rho_{32}-\rho_{31}\rho_{13}, \nonumber \\
 \alpha_{3}&=& -\rho_{11}\rho_{22}\rho_{33}-\rho_{12}\rho_{23}\rho_{31}-\rho_{13}\rho_{32}\rho_{21} +\rho_{11}\rho_{23}\rho_{32}+\rho_{22}\rho_{31}\rho_{13}+\rho_{33}\rho_{12}\rho_{21}.
 \end{eqnarray}
 It is valuable to declare that since the parameter $\alpha_{1}$ in (\ref{vzal}) specifies the trace of density matrix with minus sign, thus the exact value of this parameter is clearly equal to $-1$. Concerning equations (\ref{sff})--(\ref{vzal}), we are able to determine the variation of the entropy of the atom (or the field) with time. In addition, the DEM between the atom and field can also be specified. The pure state $|\psi(t)\rangle$ of a bipartite (atom-field) system is entangled if and only if the reduced density operators for the subsystems describe mixed states ($S_{A(F)}\neq 0$). Also, the subsystems are disentangled (the system of atom-field is separable) if DEM in equation (\ref{sff}) tends to zero ($S_{A(F)}=0$). \\
 Without loss of generality, we consider $\lambda_{1}=\lambda_{2}=\lambda$ in all of our numerical calculations in the remainder of the paper.
 Accordingly, we can plot all required quantities as a function of scaled time $\tau=\lambda t$. Also, two modes of the cavity field are considered to be initially in a standard coherent states with the mean photon numbers $|\alpha_{1}|^{2}$ and $|\alpha_{2}|^{2}$ as
 \begin{eqnarray}\label{vqf}
 |\alpha_{i} \rangle &=& \sum_{n_{i}=0}^{+\infty}q_{n_{i}} |n_{i} \rangle, \;\;\; q_{n_{i}} = \exp\left(-\frac{|\alpha_{i}|^{2}}{2}\right)\frac{\alpha_{i}^{n_{i}}}{\sqrt{n_{i}!}},\;\;\;i=1,2.
 \end{eqnarray}
 Figure 2 shows the evolution of the field entropy against the scaled time $\tau$ for initial mean number of photons fixed at
 $|\alpha_{1}|^{2} = 10=|\alpha_{2}|^{2}$.  The left plots concern with the absence of the intensity-dependent coupling, i.e. $f_{i} (n_{i}) = 1$  and in the right plots  the intensity-dependent coupling  with $f_{i} (n_{i})  = \sqrt{n_{i}}$ is considered. The physical interest in  this nonlinearity function and its associated coherent state is arisen  naturally from the Hamiltonian illustrating the interaction with intensity-dependent coupling between a two-level atom and a radiation field \cite{fn}.
 In figure 2(a) the Kerr effect is eliminated  ($\chi = 0$) and the exact resonant case is assumed ($\Delta_{2} = \Delta_{3} =0$).
 Figure 2(b) shows the effect of the Kerr medium ($\chi = 0.4 \lambda, g_{i}(n_{i})=1$) in exact resonance condition and in figure 2(c) the effect of deformed Kerr medium ($\chi = 0.4 \lambda$), which is distinguished from Kerr medium by nonlinearity function $ g_{i}(n_{i})  = 1/\sqrt{n_{i}}$, in the absence of the detuning is examined. The effect of the detuning parameters ($\Delta_{2} =7 \lambda,  \Delta_{3} =15 \lambda$) in the absence of the Kerr medium ($\chi = 0$) has been shown in figure 2(d).
 Also, the influence of these parameters in the presence of Kerr medium and deformed Kerr medium has been depicted in figures 2(e) and 2(f), respectively. It is useful to notice that the function  $ g_{i}(n_{i})  = 1/\sqrt{n_{i}}$, which has been introduced by Man'ko {\it et al} \cite{manko} (and the corresponding coherent states have been named by Sudarshan as harmonious states \cite{harmonious}), is an applied nonlinearity function which has been frequently utilized in recent literature \cite{example-harmonious}.

 In detail, from the left plot of figure 2(a), whereas the intensity-dependent coupling, the Kerr effect and the detuning parameters are neglected, a random behaviour for the time evolution of the field entropy is observed. By entering the effect of intensity-dependent coupling (right plot), the temporal behavior of the field entropy is oscillatory and the maximum amount of the DEM is increased. A random behaviour for both plots of figure 2(b) is seen in the presence of ordinary Kerr effect and in resonance. Figure 2(c) refers to the effect of $` `$deformed$"$ Kerr medium which is distinguishable from Kerr medium by the nonlinearity function $g_{i}(n_{i}) = 1/\sqrt{n_{i}}$. It is obviously beheld that the time evolution of field entropy for figure 2(c) and 2(a) is qualitatively identical. The left plot of figure 2(c) shows that the effect of deformed Kerr medium causes the maximum values of DEM to be enhanced, when it is compared with the left plot of figure 2(a). Also, if we compare figures 2(b) and 2(c), it is understood that the deformed Kerr medium increases the DEM. Specifically, focusing on the right plots, it is demonstrated that the influence of deformed Kerr medium raises up the maximum amount of the field entropy. The effect of the detuning is studied in figure 2(d). According to this figure, one can observe that in the left (right) plot of figure 2(d), the time evolution of the field entropy has an oscillatory behaviour which rapidly varies against scaled time. Although, the presence of intensity-dependent coupling (right plot of figure 2(d)) increases the amount of entanglement by a relative amount of nearly $0.3$. The simultaneous study of the influences of the detuning parameters and Kerr medium (deformed Kerr medium) has been shown in figure 2(e) (figure 2(f)). By comparison between figures 2(c) and 2(e) one can deduce that the detuning decreases the maximum amount of entanglement. In addition, comparing the right plots of these figures shows that the presence of the detuning parameters destroys the oscillatory behaviour of the field entropy in the intensity-dependent coupling regime. Finally, the influence of deformed Kerr medium in nonresonance conditions is studied in figure 2(f). Indeed, in the right plot of figure 2(f), we have entered all parameters; i.e. detuning and deformed Kerr medium in the presence of intensity-dependent coupling. From the left plots of figures 2(d) and 2(f), it is evident that opposite the ordinary Kerr medium in which the amount of the DEM has been descended by the presence of detuning parameters, they have no remarkable effect on the maximum values of the DEM, when the deformed Kerr medium is present.

 In summary, adding our above results, comparing the left plots of figure 2 (in which the atom-field couplings are constant) with the right ones shows that the presence of intensity-dependent coupling increases the maximum amount of DEM between the atom and the field. By focusing on the effect of nonlinearities associated with Kerr and deformed Kerr medium, it is found that the existence of Kerr medium and specially deformed Kerr medium (by selecting the special nonlinearity functions $g_{i}(n_{i}) = 1/\sqrt{n_{i}}$) may improve the amount of DEM, while the detuning parameters reduce this quantity.

 \section{Position-momentum entropic uncertainty relation}
 Based on the Heisenberg uncertainty principle \cite{heisenberg}, any  two noncommuting observables cannot be simultaneously measured with arbitrary precision. For example, the uncertainty for position and momentum read as $\Delta \mathbf{x} \Delta \mathbf{p} \geq 0.5$. Following Shannon's ideas, one can supply an alternative mathematical formulation of the uncertainty principle by the inequality $\delta \mathbf{x}\delta \mathbf{p} \geq \pi e$ \cite{ERspringer}, where $\delta \mathbf{x}$ and $\delta \mathbf{p}$ are defined as the exponential of Shannon entropies associated with the probability distributions for $\mathbf{x}$ and $\mathbf{p}$ as given below \cite{honarasa1,Orlowski}
 \begin{eqnarray}\label{shen}
 \delta \mathbf{x} &=&\exp (E_{x})=\exp\left(-\int \limits_{-\infty}^{+\infty}\langle x|\hat{\rho}_{F}|x\rangle \ln \langle x|\hat{\rho}_{F}|x\rangle dx \right), \nonumber \\
 \delta \mathbf{p} &=&\exp (E_{p})=\exp\left(-\int \limits_{-\infty}^{+\infty}\langle p|\hat{\rho}_{F}|p\rangle \ln \langle p|\hat{\rho}_{F}|p\rangle dp \right),
 \end{eqnarray}
 where $\hat{\rho}_{F}=\mathrm{Tr}_{A}\left(  |\psi(t) \rangle \langle \psi(t) |   \right)$ and the density matrix element may be determined as follows
 \begin{eqnarray}\label{xrx}
 \langle x|\hat{\rho}_{F}|x\rangle &=& \left| \sum_{n_{1}=0}^{+\infty} \sum_{n_{2}=0}^{+\infty} q_{n_{1}} q_{n_{2}} A(n_{1},n_{2},t) \langle x|n_{1}\rangle\right|^{2}
 +\left| \sum_{n_{1}=0}^{+\infty} \sum_{n_{2}=0}^{+\infty} q_{n_{1}} q_{n_{2}} B(n_{1}+1,n_{2},t) \langle x|n_{1}+1\rangle\right|^{2} \nonumber \\
 &+& \left| \sum_{n_{1}=0}^{+\infty} \sum_{n_{2}=0}^{+\infty} q_{n_{1}} q_{n_{2}} C(n_{1},n_{2}+1,t) \langle x|n_{1}\rangle\right|^{2},
 \end{eqnarray}
 in which
 \begin{eqnarray}\label{x-n}
 \langle x|n_{1}\rangle=\left[ \frac{\exp\left(-x^{2}\right)}{\sqrt{\pi}2^{n_{1}}n_{1}!} \right]^{1/2} H_{n_{1}}(x),
 \end{eqnarray}
 with $H_{n_{1}}(x)$ as the Hermite polynomials. Also, it is convenient to see that we restricted our calculation only for field mode 1, as can be seen in relations (\ref{xrx}) and (\ref{x-n}).\\
 To analyze the position-momentum entropic uncertainty relation or entropy squeezing of the atom in motion, we introduce two (normalized) quantities as follows:
 \begin{eqnarray}\label{quan}
 \mathbf{E}_{X}(t)&=&(\pi e)^{-1/2}\exp(E_{x}(t))-1, \nonumber \\
 \mathbf{E}_{P}(t)&=&(\pi e)^{-1/2}\exp(E_{p}(t))-1.
 \end{eqnarray}
 When $-1<\mathbf{E}_{X}(t)<0$ ($-1<\mathbf{E}_{P}(t)<0$), the position (momentum) i.e., $X(P)$ component of the field entropy is said to be squeezed.\\
 Presented results in figure 3 show the time evolution of entropy squeezing versus the scaled time $\tau$ for the initial mean number of photons fixed  at $|\alpha_{1}|^{2} = 10 = |\alpha_{2}|^{2}$. The left (right) plots again correspond to the case $f_{i}(n_{i}) = 1$ ($f_{i} (n_{i})  =  \sqrt{n_{i}}$).
 Figure 3(a) deals with the effect of intensity-dependent coupling in the absence of other effects (no Kerr medium and no resonance condition). Figure 3(b) refers to the effect of Kerr medium  ($\chi = 0.4 \lambda, g_{i}(n_{i})=1$) in the resonance condition and in figure 3(c) the effect of deformed Kerr medium ($\chi = 0.4 \lambda, g_{i} (n_{i})  = 1/\sqrt{n_{i}}$) in the absence of the detuning is studied. The effect of the detuning parameters ($\Delta_{2} =7 \lambda,  \Delta_{3} =15 \lambda$) in the absence of the Kerr medium ($\chi = 0$) is presented in figure 3(d). Also, the influence of the detuning together with Kerr medium and deformed Kerr medium has been exhibited in figures 3(e) and 3(f), respectively.

 In detail, as is seen from the left plot of figure 3(a), whereas Kerr effect and the detuning have been disregarded, entropy squeezing parameter possesses negative values almost for all times and this occurrence implies to the nonclassical behavior of the state vector of the whole system. If we compare the left and right plots (where intensity-dependent coupling is entered) of figure 3(a), it is seen that,  in the presence of this nonlinearity, entropy squeezing disappears as the time proceeds. The left plot of figure 3(b) showing the effect of Kerr medium, indicates that the entropy squeezing gets it minimum value at all times. By entering the effect of intensity-dependent coupling, it is observed that the entropy squeezing still remains constant at $-1$, which corresponds to minimum (maximum) amount of entropy squeezing parameter (nonclassicality). This situation is different when we entered the deformed Kerr medium effect (figure 3(c)). Comparing the left plot of figures 3(a) and 3(c) indicates that by entering the effect of deformed Kerr medium, which is different from the Kerr medium by entering the function $g_{i}(n_{i})=1/\sqrt{n_{i}}$, the amount of entropy squeezing may be changed and generally is reduced. Even, for some finite intervals of time, this nonclassical behavior is disappeared.
 The same treatment can be performed for the left plot of figure 3(d), where the effect of detuning parameters is examined. From this figure, one can observe that the amount of entropy squeezing gets positive and negative values, although the nonclassicality behavior is appeared in most of the time. Both of the left and right plots of figures 3(e) which are similar to their corresponding plots of figure 3(b), show that despite adding the effect of detuning, the value of entropy squeezing remains minimum ($-1$) at all times. Figure 3(f) shows that in the presence of deformed Kerr medium, the detuning parameters have no significant role in the amount of the negativity of entropy squeezing and it is clear that the time evolution of the entropy squeezing (figures 3(d) and 3(f)) are alike.

 So, we may conclude that, while the Kerr medium has a direct role in revealing the squeezing in entropy of the state vector of the system, the deformed Kerr effect and also the detuning parameters reduce the strength of entropy squeezing, as a nonclassicality sign. Also, it may be understood from the right plots of figure 3 that, the effect of intensity-dependent coupling (which characterized by the nonlinearity function $f_{i}(n_{i})=\sqrt{n_{i}}$) descends the amount of negativity of entropy squeezing and so decreases the nonclassicality of the system state.


 \section{Summary and conclusion}\label{summary}
 In this paper, we have discussed the nonlinear interaction between a three-level atom (which was considered to be in a $\Lambda$-configuration) and a two-mode cavity field in the presence of cross-Kerr medium and its deformed counterpart (which are distinguished from each other by the nonlinearity function $g_{i}(n_{i})$), intensity-dependent atom-field coupling (which is characterized by the nonlinearity function $f_{i}(n_{i})$) and the detuning parameters. Next, after obtaining the explicit analytical, even though complicated form of the state vector of the bipartite (atom-field) system in a general manner, quantum entanglement between the subsystems has been evaluated computationally, by using the approach of von Neumann entropy. Also, considering the Shannon's idea, entropy squeezing or position-momentum entropic uncertainty relation of the state vector of the entire system has been examined, numerically. In each case, we individually studied the effects of $` `$intensity-dependent coupling$"$ (entered by $f_{i}(n_{i})=\sqrt{n_{i}}$), $` `$Kerr medium$"$,  $` `$deformed Kerr medium$"$ (entered by $g_{i}(n_{i})=1/\sqrt{n_{i}}$) and $` `$detuning parameters$"$ on the physical quantities consist of quantum entropy of the field and position-momentum entropic uncertainty relation (entropy squeezing).

 The numerical results of the field entropy showed that the existence of Kerr medium (and also deformed Kerr medium) may enhance the maximum amount of entanglement between the atom and the field, while the detuning parameters and intensity-dependent coupling reduce the DEM between subsystems. Also, the numerical results of the entropy squeezing (position-momentum entropic uncertainty relation) indicate that Kerr medium has a significant role in increasing the negativity of entropy squeezing in position component and so the nonclassicality of the state. Although, the deformed Kerr effect and also the detuning parameters reduce the strength of this nonclassicality sign of the state vector of the system. In addition, from the right plots of figure 3 where the effect of intensity--dependent coupling has been studied, it is observed that this effect (of course, by considering the chosen nonlinearity function $f_{i}(n_{i}) = \sqrt{n_{i}}$) descends the amount of negativity of entropy squeezing.

 Finally, we would like to emphasize the generality of our obtained formalism in the sense that it may be used for any physical system, either any nonlinear oscillator algebra with arbitrary nonlinearity function, or any solvable quantum system with known discrete energy spectrum $e_{n}$ \cite{honarasa2}.

 \begin{flushleft}
 {\bf Acknowledgements}\\
 \end{flushleft}
 The authors would like to thank the referees who have made valuable and careful comments, which improved the paper considerably. Also, one of the authors (MJF) is indebted to Ali Shakiba for his valuable help in the numerical results.



\begin{thebibliography}{99}

 \bibitem{JCM} E. T. Jaynes, F. W. Cummings, $` `$Comparison of quantum and semiclassical radiation theories with application to the beam maser$"$ Proc. IEEE. {\bf 51}, 89--109 (1963); \\
               F. W. Cummings, $` $Stimulated emission of radiation in a single mode$"$ Phy. Rev. {\bf 140}, A1051--A1056 (1965).

 \bibitem{suk} B. Buck, C. V. Sukumar, $` `$Exactly soluble model of atom-phonon coupling showing periodic decay and revival$"$ Phys. Lett. A {\bf 81}, 132--135 (1981); \\
                C. V. Sukumar, B. Buck, $` `$Multi-phonon generalisation of the Jaynes-Cummings model$"$ Phys. Lett. A {\bf 83}, 211--213 (1981).

 \bibitem{buzek} V. Bu$\check{\mathrm{z}}$ek, $` `$Jaynes-Cummings model with intensity-dependent coupling interacting with Holstein-Primakoff SU(1, 1) coherent state$"$ Phys. Rev. A {\bf 39}, 3196--3199 (1989).

 \bibitem{selective} F. An-fu, W. Zhi-wei, $` `$Phase, coherence properties, and the numerical analysis of the field in the nonresonant Jaynes-Cummings model$"$ Phys. Rev. A {\bf 49}, 1509--1512 (1994); \\
                     R. H. Xie, G. O. Xu, D. H. Liu, $` `$Numerical study of non-classical effects and the effect of virtual photon fields in the Jaynes-Cummings model$"$ Phys. Lett. A {\bf 202}, 28--33 (1995); \\
                     V. I. Koroli, V. V. Zalamai, $` `$Dynamics of a laser-cooled and trapped radiator interacting with the Holstein-Primakoff  SU(1,1) coherent state$"$ J. Phys. B: At. Mol. Opt. Phys. {\bf 42}, 035505 (2009);\\
                     M. K. Tavassoly, F. Yadollahi, $` `$Dynamics of states in the nonlinear interaction regime between a three-level atom and generalized coherent states and their non-classical features$"$ Int. J. Mod. Phys. B {\bf 26}, 1250027 (2012).


 \bibitem{mao} M. F. Fang, H. E. Liu, $` `$Properties of entropy and phase of the field in the two-photon Jaynes-Cummings model with an added Kerr medium$"$ Phys. Lett. A {\bf 200}, 250--256 (1995).

 \bibitem{kazakov} A. Y. Kazakov, $` `$Modified Jaynes-Cummings model: Interaction of the two-level atom with two modes$"$ Phys. Lett. A {\bf 206}, 229--234 (1995).

 \bibitem{crn} J. Crnugelj, M. Martinis, V. Mikuta-Martinis, $` `$Properties of a deformed Jaynes-Cummings model$"$ Phys. Rev. A {\bf 50}, 1785--1791 (1994).

 \bibitem{wahab} N. H. Abdel-Wahab, M. F. Mourad, $` `$On the interaction between two two-level atoms and a two mode cavity field in the presence of detuning and cross-Kerr nonlinearity$"$ Phys. Scr. {\bf 84}, 015401 (2011).

 \bibitem{zait} R. A. Zait, $` `$Nonclassical statistical properties of a three-level atom interacting with a single-mode field in a Kerr medium with intensity dependent coupling$"$ Phys. Lett. A {\bf 319}, 461--474 (2003).

 \bibitem{us} M. J. Faghihi, M. K. Tavassoly, $` `$Dynamics of entropy and nonclassical properties of the state of a $\Lambda$-type three-level atom interacting with a single-mode cavity field with intensity-dependent coupling in a Kerr medium$"$ J. Phys. B: At. Mol. Opt. Phys. {\bf 45}, 035502 (2012).

 \bibitem{guo} J. L. Guo, Y. B. Sun, Z. D. Li, $` `$Entropy exchange and entanglement in Jaynes-Cummings model with Kerr-like medium and intensity-depend coupling$"$ Opt. Commun. {\bf 284}, 896--901 (2011).

 \bibitem{rec1} S. Cordero, J. R$\acute{\mathrm{e}}$camier, $``$Algebraic treatment of the time-dependent Jaynes-Cummings Hamiltonian including nonlinear terms$"$ J. Phys. A: Math. Theor. {\bf 45}, 385303 (2012).

 \bibitem{rec2} O. de. los S. -S$\acute{\mathrm{a}}$nchez, J. R$\acute{\mathrm{e}}$camier, $``$The $f$-deformed Jaynes-Cummings model and its nonlinear coherent states$"$ J. Phys. B: At. Mol. Opt. Phys. {\bf 45}, 015502 (2012).


 \bibitem{aty} M. Abdel-Aty, A. S. F. Obada, $` `$Engineering entanglement of a general three-level system interacting with a correlated two-mode nonlinear coherent state$"$ Eur. Phys. J. D {\bf 23}, 155--165 (2003).

 \bibitem {zubairy} M. O. Scully, M. S. Zubairy, {\it Quantum Optics} (Cambridge: Cambridge University Press, 2001).

 \bibitem{newhonarasa} G. R. Honarasa, and M. K. Tavassoly, $` `$Generalized deformed Kerr states and their physical properties$"$ Phys. Scr. {\bf 86}, 035401 (2012).

 \bibitem{kardan} N. C. Lindsay, {\it A Concrete Introduction to Higher Algebra} (3rd edn, Berlin: Springer, 2008).

 \bibitem{benenti}  G. Benenti, G. Casati, G. Strini, {\it Principles of Quantum Computation and Information, Vols I and II} (Singapore: World Scientific, 2007).

 \bibitem{chuang}  M. A. Nielsen, I. L. Chuang, {\it Quantum Computation and Quantum Information} (Cambridge: Cambridge University Press, 2000).

 \bibitem{pk1} S. J. D. Phoenix, P. L. Knight, $` `$Establishment of an entangled atom-field state in the Jaynes-Cummings model$"$ Phys. Rev. A {\bf 44}, 6023--6029 (1991).

 \bibitem{araki} M. Araki, E. Leib, $` `$Entropy inequalities$"$ Commun. Math. Phys. {\bf 18}, 160--170 (1970).

 \bibitem{phoenix} S. M. Barnett, S. J. D. Phoenix, $` `$Information theory, squeezing, and quantum correlations$"$ Phys. Rev. A {\bf 44}, 535--545 (1991).

 \bibitem{pk3} S. J. D. Phoenix, P. L. Knight, $` `$Periodicity, phase, and entropy in models of two-photon resonance$"$ J. Opt. Soc. Am. B {\bf 7}, 116--124 (1990).

  \bibitem{fn} G. S. Agarwal, S. Singh, $``$Effect of pump fluctuations on line shapes in coherent anti-Stokes Raman scattering$"$ Phys. Rev. A {\bf 25}, 3195--3205 (1982); \\
                     C. V. Sukumar, B. Buck, $``$Solution of the Heisenberg equations for an atom interacting with radiation$"$ J. Phys. A: Math. Gen. {\bf 17}, 877 (1984); \\
                     F. Eftekhari, M. K. Tavassoly, $``$On a general formalism of nonlinear charge coherent states, their quantum statistics and nonclassical properties$"$ Int. J. Mod. Phys. A {\bf 25}, 3481--3504 (2010); \\
                     O. Safaeian, M. K. Tavassoly, $``$Deformed photon-added nonlinear coherent states and their non-classical properties$"$ J. Phys. A: Math. Theor. {\bf 44}, 225301 (2011).

 \bibitem{manko} V. I. Man'ko, G. Marmo, E. C. G. Sudarshan, F. Zaccaria, $``$$f$-oscillators and nonlinear coherent states$"$ Phys. Scr. {\bf 55}, 528--541 (1997).

 \bibitem{harmonious} E. C. G. Sudarshan, $``$Diagonal harmonious state representations$"$ Int. J. Theor. Phys. {\bf 32}, 1069--1076 (1993).

 \bibitem{example-harmonious} M. K. Tavassoly, $``$New nonlinear coherent states associated with inverse bosonic and $f$-deformed ladder operators$"$ J. Phys. A: Math. Theor. {\bf 41}, 285305 (2008);\\
     %
     M. K. Tavassoly, $``$On the non-classicality features of new classes of nonlinear coherent states$"$ Opt. Commun. {\bf 283}, 5081--5091 (2010);\\
     %
     E. Piroozi, M. K. Tavassoly, $``$Nonlinear semi-coherent states, their nonclassical features and phase properties$"$ J. Phys. A: Math. Theor. {\bf 45}, 135301 (2012).

 \bibitem{heisenberg} W. Heisenberg, $` `$The actual content of quantum theoretical kinematics and mechanics$"$ Z. Phys. {\bf 43}, 172--198 (1927).

 \bibitem{ERspringer} I. Bia{\l}ynicki-Birula, J. Mycielski, $` `$Uncertainty relations for information entropy in wave mechanics$"$ Commun. Math. Phys. {\bf 44}, 129--132 (1975).

 \bibitem{honarasa1} G. R. Honarasa, M. K. Tavassoly, M. Hatami, $` `$Number-phase entropic uncertainty relations and Wigner functions for solvable quantum systems with discrete spectra$"$ Phys. Lett. A {\bf 373}, 3931--3936 (2009).

 \bibitem{Orlowski} A. Or{\l}owski, $` `$Information entropy and squeezing of quantum fluctuations$"$ Phys. Rev. A {\bf 56}, 2545--2548 (1997).

 \bibitem{honarasa2} G. R. Honarasa, M. K. Tavassoly, M. Hatami, $` `$Quantum phase properties associated to solvable quantum systems using the nonlinear coherent states approach$"$ Opt. Commun. {\bf 282}, 2192--2198 (2009); \\
     V. I. Man'ko, G. Marmo, F. Zaccaria, $` `$Moyal and tomographic probability representations for $f$-oscillator quantum states$"$ Phys. Scr. {\bf 81}, 045004 (2010); \\
                                     M. J. Faghihi, M. K. Tavassoly, $` `$Nonlinear quantum optical springs and their nonclassical properties$"$ Commun. Theor. Phys. {\bf 56}, 327--332 (2011).

\end{thebibliography}
 \end{document}